\documentclass{article}
\usepackage{spconf,amsmath,graphicx,hyperref}
\usepackage{xcolor, soul}
\usepackage{booktabs}
\usepackage{subcaption}
\usepackage{setspace}
\usepackage{multirow}
\usepackage{fancyhdr}
\usepackage[dvipsnames]{xcolor}
\usepackage[colorinlistoftodos,obeyFinal]{todonotes} 

\title{Scaling Ambiguity: Augmenting Human Annotation in Speech Emotion Recognition with Audio-Language Models}

\name{Wenda Zhang$^{1}$ \qquad Hongyu Jin$^{1,\ast}$\thanks{$^{\ast}$Equal contribution} \qquad Siyi Wang$^{1,\ast}$ \qquad Zhiqiang Wei$^{2}$ \qquad Ting Dang$^{1}$}

\address{$^{1}$ University of Melbourne, Melbourne, Australia\\ 
$^{2}$ Xi'an Jiaotong University, Xi'an, China\\
}

\begin{document}
\ninept 

\maketitle

\thispagestyle{fancy}
\fancyhf{}
\renewcommand{\headrulewidth}{0pt}
\fancyfoot[L]{\footnotesize \copyright\ 2026 IEEE. Personal use of this material is permitted.  Permission from IEEE must be obtained for all other uses, in any current or future media, including reprinting/republishing this material for advertising or promotional purposes, creating new collective works, for resale or redistribution to servers or lists, or reuse of any copyrighted component of this work in other works.}
\setlength{\footskip}{23pt}

\maketitle

\begin{abstract}
Speech Emotion Recognition models typically use single categorical labels, overlooking the inherent ambiguity of human emotions. Ambiguous Emotion Recognition addresses this by representing emotions as probability distributions, but progress is limited by unreliable ground-truth distributions inferred from sparse human annotations. This paper explores whether Large Audio-Language Models (ALMs) can mitigate the annotation bottleneck by generating high-quality synthetic annotations. We introduce a framework leveraging ALMs to create Synthetic Perceptual Proxies, augmenting human annotations to improve ground-truth distribution reliability. We validate these proxies through statistical analysis of their alignment with human distributions and evaluate their impact by fine-tuning ALMs with the augmented emotion distributions. Furthermore, to address class imbalance and enable unbiased evaluation, we propose DiME-Aug, a \underline{Di}stribution-aware \underline{M}ultimodal \underline{E}motion \underline{Aug}mentation strategy. Experiments on IEMOCAP and MSP-Podcast show that synthetic annotations enhance emotion distribution, especially in low-ambiguity regions where annotation agreement is high. However, benefits diminish for highly ambiguous emotions with greater human disagreement. This work provides the first evidence that ALMs could address annotation scarcity in ambiguous emotion recognition, but highlights the need for more advanced prompting or generation strategies to handle highly ambiguous cases.
\end{abstract}


\vspace{-5pt}
\section{Introduction}
\label{sec:intro}
\vspace{-5pt}

Speech emotion recognition (SER) has gained increasing attention in areas such as conversational agents \cite{zhou2020design, lee2020study} and mental health monitoring \cite{chakravarthi2022eeg}. A fundamental challenge is that human emotions are inherently ambiguous and subjective, which can not be adequately represented with a single categorical label \cite{wani2021comprehensive}. To better account for this inherent ambiguity, recent work has shifted toward modeling emotions as distributions\cite{wu2024emotion}, leading to the emerging field of Ambiguity Emotion Recognition (AER) \cite{hong2025aer, halim2025token}.

Despite this progress, the effectiveness of AER is fundamentally constrained by the quality of its ground-truth distributions, which are typically inferred from sparse human annotators often comprising 3 to 5 ratings per utterance~\cite{iemocap,msp}. This sparsity yields coarsely estimated distributions that fail to capture the full spectrum of emotional ambiguity. Consequently, models trained on this data are limited in their capability to learn the true underlying emotional ambiguity, which impairs their performance and generalization.

Recent advancements in Large Language Models (LLMs) and Large Audio-Language Models (ALMs) have opened new possibilities, including AER. A few studies have begun to explore their potential for generating synthetic annotations to augment sparse human labels, thereby refining ground truth estimation~\cite{textbaseannotation}. Niu et al.\cite{LLMgoodemotionannotation} showed that emotion labels generated by LLM align closely with human annotations and are often preferred by evaluators. In contrast, Feng et al.\cite{cannotsolelydependonLLM} found that fully relying on LLM-generated emotion annotations is unreliable, highlighting the importance of incorporating human feedback. 
These initial studies demonstrate both the promise and the limitations of synthetic annotation in emotion recognition tasks, providing an inspiring foundation for further research. However, these approaches treated emotion as a single categorical label, overlooking its inherently distributional nature and the opportunity for synthetic annotations to enrich emotion distributions. Moreover, most of these studies rely on text-based LLMs, while the emergence of ALMs may represent a promising avenue for creating superior synthetic emotion annotations by incorporating acoustic cues.

In this work, we investigate whether synthetic annotations generated by ALMs can augment sparse human annotations to enable more reliable emotion distributional representations and improve ambiguity modeling in AER. We propose two approaches to evaluate the quality and utility of synthetic annotations: (i) \textbf{statistical alignment} analysis between synthetic and human distributions, and (ii) \textbf{downstream evaluation} by fine-tuning ALMs for emotion distribution estimation on different annotation sources. 
To realize this, we design a three-component framework that supports both the generation and evaluation of synthetic emotion annotations: (i) \textbf{Synthetic Perceptual Proxies}, which leverage ALMs to generate synthetic annotations and enrich emotion distributions; (ii) \textbf{DiME-Aug}, a distribution-aware multimodal augmentation strategy that addresses class imbalance common in multimodal emotion datasets \cite{iemocap,msp}, thereby enabling a reliable and unbiased evaluation of the impact of synthetic annotations;  and (iii) \textbf{ALM Fine-tuning}, where the ALM is fine-tuned on both original and augmented emotion distributions for comparison.

Our complementary evaluations reveal that with a sufficient number of synthetic annotations, the generated emotion distributions can closely approximate those derived from human annotations. When integrated with human annotations, the augmented emotion distributions can improve the reliability of AER, with this effect more pronounced in low-ambiguity regions than in high-ambiguity regions. 
To our knowledge, this is the first work exploring synthetic annotations generated by ALMs to enhance distributional labels in AER, offering a scalable solution to annotation sparsity. This enables models to better capture emotional ambiguity, offering promises for mental health applications, HCI and conversational AI. 

\begin{figure*}[t!]
    \centering
    \includegraphics[width=1\textwidth]{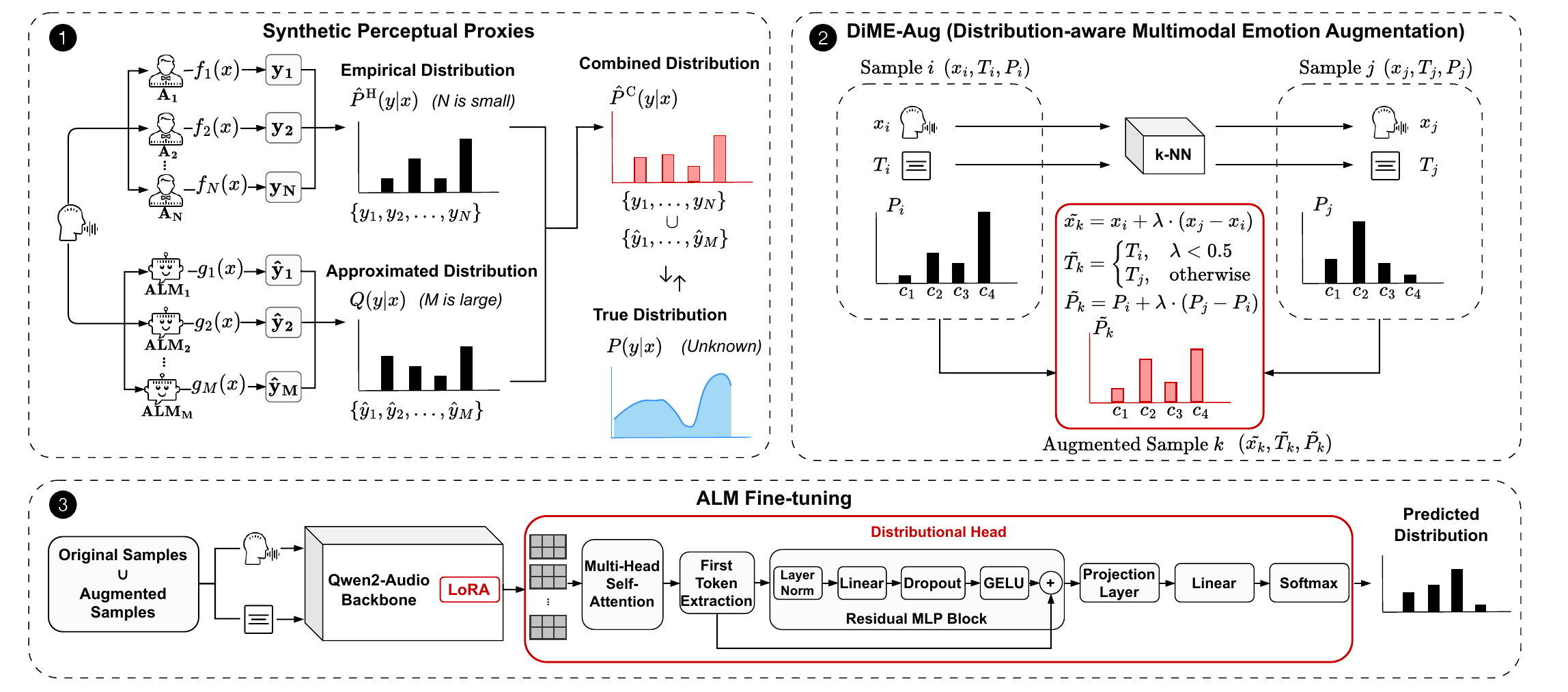}
    \vspace{-15pt}
    \caption{Overview of the proposed framework. The pipeline consists of three modules: (1) Synthetic Perceptual Proxies, which leverage ALMs to augment human annotations; (2) DiME-Aug, a distribution-aware multimodal augmentation to address class imbalance; and (3) Audio-Language Model (ALM) Fine-tuning.
    }
    \vspace{-12pt}
    \label{fig:1}
\end{figure*}
\vspace{-10pt}
\section{Methodology}
\vspace{-7pt}
\label{sec:methodology}
\subsection{Overview}

As illustrated in Fig.~\ref{fig:1}, our framework is composed of three modules:
i) \textbf{Synthetic perceptual proxies} leverages ALMs to generate synthetic emotion annotations, which are combined with human annotations to form enriched emotion distributions. ii) \textbf{DiME-Aug} augments the dataset by balancing minority classes and prepares the data for training. iii) \textbf{ALM finetuning} trains the backbone model on the augmented dataset with enriched emotion distributions. This framework thus provides a unified approach for generating, integrating, and evaluating synthetic annotations in AER.

\vspace{-7pt}
\subsection{Synthetic Perceptual Proxies}
We employ ALMs to generate synthetic annotations given the audio recordings and transcripts. 
The ALM is guided by a detailed prompt that includes: (i) A task description that prompts ALMs to infer the emotion of input speech from a predefined set of possible emotion categories; (ii) the utterance transcript and (iii) explicit instructions to analyze both vocal tone and linguistic content for a reliable emotion annotation. To simulate the diverse perspectives inherent in human emotion perception, we introduce controlled variance into the generation process through two primary mechanisms. First, we adjust the temperature parameter to control the randomness of its outputs. Second, we strategically modify the instructional prompts to guide the model to focus towards slightly different acoustic or linguistic characteristics (e.g., emphasizing vocal tone, specific word choices, or other subtle emotional cues), simulating distinct annotator personas. The prompt is shown in Table \ref{tab:prompt_template}.

By systematically varying both the prompt phrasing and the sampling temperature, we generate a large set of unique, single-label synthetic annotations for each utterance, each representing a plausible annotator's perspective. These synthetic annotations are combined with the original, limited human labels, treating each equally as a sample drawn from the underlying emotion distribution, to infer the augmented emotion distribution(Figure~\ref{fig:1}).

\begin{table}[t!]
    \centering
    \vspace{-0.2cm}
    \caption{Prompt for Synthetic Annotation Generation}
    \vspace{-0.15cm}
    \label{tab:prompt_template}
    \footnotesize
    \renewcommand{\arraystretch}{1.1}
    \begin{tabular}{p{1.5cm}|p{6.0cm}}
    \toprule
     & \textbf{Prompt Template} \\
    \midrule
    \textbf{Task \newline Description} & Listen to the audio and analyze this speech utterance. Think deeply and Select the ONE most dominant emotion present. \\
    \midrule
    \textbf{Transcription} & Transcript: "[Utterance Text]" \\
    \midrule
    \textbf{Emotion \newline Categories} & Possible emotions: [Emotion List] \newline (e.g., Angry, Happy, Sad, Neutral) \\
    \midrule
    \textbf{Instructions} & 1. Carefully analyze the emotional content in both the audio and the text. \newline 2. [Annotator persona](randomly selected from the annotator persona set). \newline 3. Think deeply and Select ONLY ONE emotion from the list that best represents the dominant emotional state. \newline 4. Respond with ONLY the emotion name, nothing else.  \\
    \midrule
    \textbf{Output} & Which single emotion best describes this utterance? Select only from: [Emotion List] \\ \midrule \midrule
    \multicolumn{2}{c}{\textbf{Annotator Persona Set}}\\ \midrule
     \multicolumn{2}{p{7.5cm}}{• Focus on both vocal tone and linguistic content. \newline • Pay special attention to the speaker's tone of voice. \newline • Consider the word choice and phrasing in the utterance. \newline • Listen for subtle emotional cues in the voice. \newline • Analyze both what is said and how it is expressed.} \\
    \bottomrule
    \end{tabular}
    \vspace{-10pt}
\end{table}

\vspace{-7pt}
\subsection{DiME-Aug (Distribution-aware Multimodal Emotion Augmentation)}

To address the class imbalance before fine-tuning our models and ensure an unbiased evaluation, we introduce DiME-Aug, a data augmentation strategy that creates samples to populate underrepresented classes for AER. Our approach adapts the mixup concept \cite{mixup1, mixup2}, which generates new data points by interpolating existing ones.

We identify the minority class based on the utterance's dominant emotion. For a given speech utterance from the minority class, we first search for its nearest neighbor in the feature space. Let an utterance $i$ be represented by a triplet containing its audio signal $x_i$, transcript $T_i$, and ground-truth emotion distribution $P_i$.
Its nearest neighbor, utterance $j$, is found using K-nearest neighbors based on the feature representations. A new augmented sample $(x_{k}, T_{k}, P_{k})$ is then generated through a weighted combination controlled by a coefficient $\lambda \in [0, 1]$ as:
\vspace{-5pt}
\begin{equation}
x_{k} = \lambda x_i + (1 - \lambda) x_j
\vspace{-2pt}
\end{equation}

Unlike existing work that augments only a single modality \cite{Yu_2025_CVPR, emmanouilidou2024domain}, our method is multimodal which requires generating a linguistically coherent transcript that remains synchronized with the augmented audio signal. To achieve this without creating nonsensical text, instead of mixing the transcripts, the new sample inherits the transcript of the dominant original utterance, either $T_i$ or $T_j$, as determined by a 0.5 threshold on the mixing coefficient $\lambda$.

Finally, the ground-truth distribution for the new sample is created by interpolating the two original distributions, simulating a blend of the two emotional states:
\vspace{-5pt}
\begin{equation}
P_{k} = \lambda P_i + (1 - \lambda) P_j
\vspace{-5pt}
\end{equation}
This interpolation serves to augment the minority class rather than precisely model mixed emotions. The multimodal augmentation process yields a new, coherent training sample that helps create a more balanced and robust training set.

\vspace{-7pt}
\subsection{Model Fine-tuning}
To evaluate whether synthetic annotations lead to improved emotion distribution, we first apply DiME-Aug to balance the dataset, and then fine-tune a backbone model using the augmented emotion distributions. Qwen2-Audio \cite{qwen2} is used as the backbone, and we introduced an additional classifier head to predict the probabilities for each of the emotion classes, inferring the predicted distributions. The distributional head consists of a multi-head self-attention mechanism, a residual multi-layer perceptron (MLP) block, a projection layer, and a linear layer with softmax for the final distributional predictions (Figure~\ref{fig:1}). The model is trained using Jensen-Shannon Divergence loss to directly optimize for distributional similarity between predicted and target emotion distributions.
\vspace{-5pt}
\section{Experimental Setup}
\label{sec:experimental}
\vspace{-7pt}

\subsection{Datasets}
Two widely used datasets were employed: IEMOCAP \cite{iemocap} and MSP-Podcast \cite{msp}. We focused on four primary emotion categories: Angry, Happy, Sad, and Neutral. IEMOCAP consists of emotional speech from dyadic actor conversations, with each utterance annotated by 3 annotators. For valid distribution inference, we selected utterances where all three annotations fall within the four chosen categories, resulting in 4,370 utterances. MSP-Podcast contains naturalistic emotional speech extracted from podcast recordings, with annotations ranging from 5 to 21 annotators per utterance. Applying the same selection criteria yielded 4,114 utterances for analysis. Both datasets are randomly split into 80/20 train/test sets.

\vspace{-7pt}
\subsection{Implementation Details}
We employ Gemini 2.5-Pro \cite{gemini25pro} to generate the synthetic annotations, owing to its strong capability in emotion understanding \cite{SERevidence1, SERevidence2}. In preliminary analysis, the model identifies negative emotional tones in 54.24\% of cases with conflicting positive text, suggesting sensitivity to acoustic cues beyond linguistic bias. To simulate variations in human perceptions, we vary the temperature parameter between 0.1 and 1.0.

Before fine-tuning, we first apply DiME-Aug to balance the dataset. The proportion of synthetic samples to add is a hyperparameter, which we optimize via a grid search over the range of 10\% to 50\% in 10\% increments. The ratio yielding the best performance is then used for the final training.

The base model for fine-tuning builds upon Qwen2-Audio-7B-Instruct \cite{qwen2}. We optimized the model using LoRA \cite{LoRA} with a rank of $r = 8$, a scaling factor of $\alpha = 16$, and a dropout rate of $0.2$, applied to the query, key, value, and output projection layers. Training utilizes mixed-precision optimization with cosine learning rate scheduling, a maximum of 50 epochs, early stopping with a patience of 8, a learning rate of $2.5e-6$, and an effective batch size of 64. Audio inputs are processed at a 16 kHz sampling rate.

To evaluate the effectiveness of the synthetic annotations to infer the emotion distributions, we compare three different settings: (1) \textit{Human-only} using only human annotations for distribution inference; (2) \textit{Synthetic-only} taking only synthetic annotations; and (3) \textit{Combined}, incorporating both human and synthetic annotations, with the number of synthetic annotations determined by the saturation analysis in \ref{sec:synthetic_analysis}.

\vspace{-7pt}
\subsection{Evaluations}
Our evaluation is twofold, assessing both the intrinsic quality of the synthetic annotations and their extrinsic impact on the downstream AER task. To intrinsically evaluate the quality of the generated annotations, we compare the emotion distributions derived by the synthetic annotations against those derived solely from human labels. We measure the similarity between these two sets of distributions using Jensen-Shannon (JS) Divergence. This serves as a sanity check to quantify the plausibility of our synthetic data. Further, we evaluate the final model performance after fine-tuning, using JS Divergence and Bhattacharyya Coefficient (BC).
\vspace{-5pt}
\section{Results}
\vspace{-5pt}
\label{sec:results}
In this section, we aim to answer three questions: (i) To what extent can synthetic annotations approximate human emotion distributions? (ii) Do synthetic annotations improve downstream performance in ambiguous emotion recognition? and (iii) How effective is DiME-Aug in mitigating class imbalance and enhancing distributional modeling?
\vspace{-7pt}
\subsection{Similarity between Human and Synthetic Annotations}
\label{sec:synthetic_analysis}

\begin{figure}[t!]
    \centering
    \begin{subfigure}[b]{0.45\linewidth}
        \centering
        \includegraphics[width=\linewidth, trim=0 10 0 0, clip]{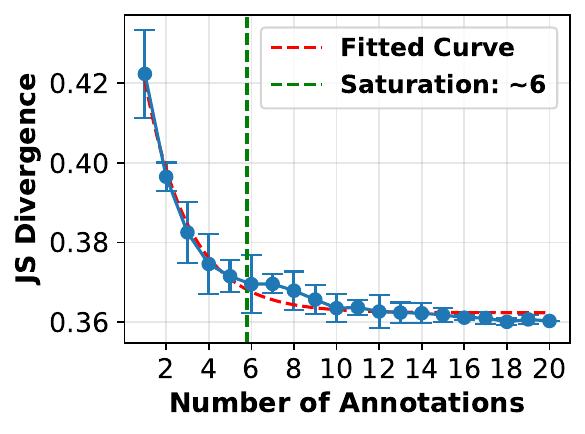}
        \caption{IEMOCAP}
        \label{fig:js-iemocap}
    \end{subfigure}
       \hspace{0.04\linewidth}
    \begin{subfigure}[b]{0.45\linewidth}
        \centering
        \includegraphics[width=\linewidth, trim=0 10 0 0, clip]{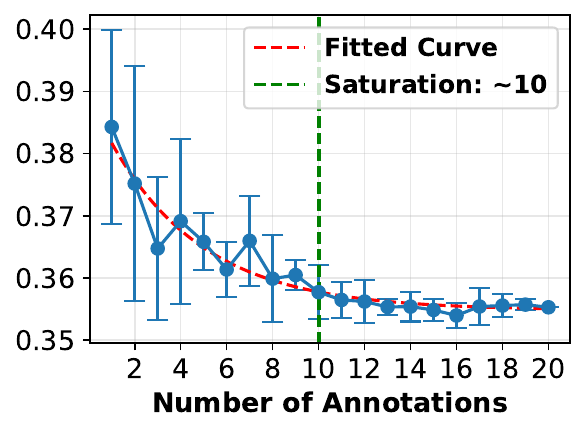}
        \caption{MSP-Podcast}
        \label{fig:js-msp}
    \end{subfigure}
    \vspace{-7pt}
    \caption{JS Divergence (lower is better) vs. number of annotations on (a) IEMOCAP and (b) MSP-Podcast datasets. The dashed lines show fitted curves and saturation points.}
    \vspace{-15pt}
    \label{fig:js-synthetic}
\end{figure}

Figure~\ref{fig:js-synthetic} plots the JS divergence between the synthetic and human-derived distributions as a function of the number of synthetic annotations. For both IEMOCAP and MSP-Podcast, JS divergence consistently decreases as more synthetic annotations are added, indicating closer alignment with human distributions and improved distributional consistency. Convergence occurs with approximately six synthetic annotations for IEMOCAP and ten for MSP-Podcast. This gap reflects dataset characteristics: MSP-Podcast provides more human annotators per utterance (5-21 in MSP-Podcast vs. 3 in IEMOCAP), resulting in richer and more diverse distributional patterns. Consequently, a larger number of synthetic annotations is required to approximate the complexity of these naturalistic distributions. While they become more aligned as the number of annotators increases, they are not exactly matching with human annotators, with JS values around 0.35. This is not necessarily a limitation, as the human distributions are not a perfect ground truth; they are themselves estimates from sparse and noisy labels. The goal is plausible augmentation, not perfect replication of an imperfect reference.

\begin{table}[t]
\centering
\caption{Performance of AER. H, S, and C denote human, synthetic, and combined annotations; 
($\uparrow$)/($\downarrow$) indicate that higher/lower values are better, respectively. w/ Aug. and w/o Aug. refer to training with or without the DiME-Aug strategy. \textbf{Bold} numbers indicate the best performance.}
\vspace{-5pt}
\label{tab:dime-results}
\footnotesize
\renewcommand{\arraystretch}{1.1}
\begin{tabular*}{0.9\columnwidth}{@{\extracolsep{\fill}}c|ccc|ccc@{}}
\toprule
\multicolumn{1}{c}{} & \multicolumn{6}{c}{IEMOCAP} \\
\midrule
& \multicolumn{3}{c|}{JS $\downarrow$} & \multicolumn{3}{c}{BC $\uparrow$} \\
Annot. & H & S & C & H & S & C \\
\midrule
w/ Aug.  & \textbf{0.302} & 0.431 & 0.325 & \textbf{0.724} & 0.607 & 0.715 \\
w/o Aug. & 0.351 & 0.480 & 0.409 & 0.679 & 0.568 & 0.642 \\
\midrule
\multicolumn{1}{c}{} & \multicolumn{6}{c}{MSP-Podcast} \\
\midrule
& \multicolumn{3}{c|}{JS $\downarrow$} & \multicolumn{3}{c}{BC $\uparrow$} \\
Annot. & H & S & C & H & S & C \\
\midrule
w/ Aug.  & 0.307 & 0.373 & \textbf{0.274} & 0.719 & 0.660 & \textbf{0.757} \\
w/o Aug. & 0.371 & 0.321 & 0.383 & 0.663 & 0.711 & 0.665 \\
\bottomrule
\end{tabular*}
\end{table}

\vspace{-7pt}
\subsection{Impact of Annotation Sources on AER}
\begin{table}[t]
    \centering
    \caption{Annotation Statistics. H, S and C represent human, synthetic and combined annotations.}
    \vspace{-5pt}
    \label{tab:train-stats}
    \footnotesize
    \renewcommand{\arraystretch}{1.1}
    \begin{tabular*}{0.9\columnwidth}{@{\extracolsep{\fill}}c|ccc|ccc@{}}
    \toprule
    \multirow{2}{*}{Metric} & \multicolumn{3}{c|}{IEMOCAP} & \multicolumn{3}{c}{MSP-Podcast} \\
    & H & S & C & H & S & C \\
    \midrule
    F-Kappa & 0.542 & 0.803 & 0.563 & 0.704 & 0.778 & 0.520 \\
    Entropy & 0.431 & 0.255 & 0.633 & 0.309 & 0.228 & 0.552 \\
    \bottomrule
    \end{tabular*}
\end{table}

\begin{figure}[t!]
    \centering
    \begin{subfigure}[b]{0.45\linewidth}
        \centering
        \includegraphics[width=\linewidth, trim=0 6 0 0, clip]{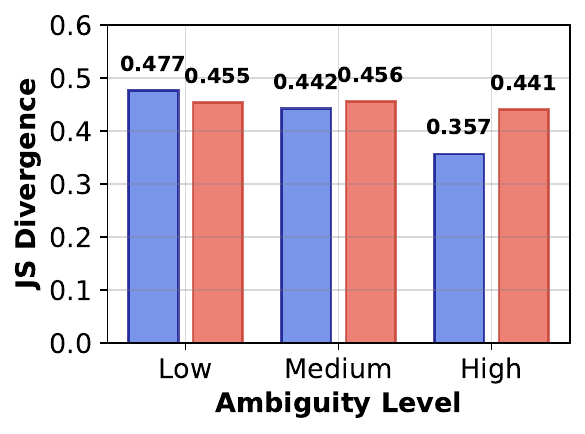}
        \caption{IEMOCAP}
        \label{fig:ambiguity-iemocap}
    \end{subfigure}
       \hspace{0.04\linewidth}
    \begin{subfigure}[b]{0.45\linewidth}
        \centering
        \includegraphics[width=\linewidth, trim=0 6 0 0, clip]{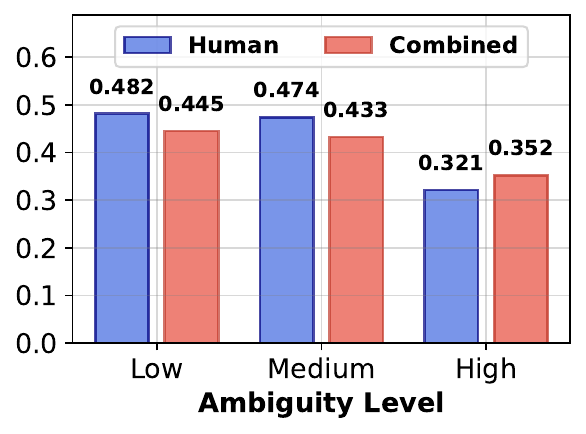}
        \caption{MSP-Podcast}
        \label{fig:ambiguity-msp}
    \end{subfigure}
    \vspace{-5pt}
    \caption{JS divergence (lower is better) vs. ambiguity level on (a) IEMOCAP and (b) MSP-Podcast datasets. Samples are divided into three groups based on the human annotation's Shannon Entropy: Low, Medium, and High.}
    \vspace{-10pt}
    \label{fig:ambiguity-analysis}
\end{figure}

We evaluate the ALM fine-tuned with DiME-Aug on three different annotation sources for ambiguous SER, with results summarized in Table~\ref{tab:dime-results}. We found that models trained on combined annotations achieve performance comparable or superior to human-only training, indicating that synthetic labels can complement human annotations when integrated appropriately. However, synthetic-only training consistently yields the weakest results, underscoring that synthetic labels alone are insufficient for accurate distributional modeling.

The datasets display different trends: MSP-Podcast shows the largest gains from combined annotations, while IEMOCAP performs best with human-only labels. To gain a deeper insights, we compute Fleiss' Kappa (F-Kappa) which measures inter-annotator agreement and Shannon Entropy quantifying distributional ambiguity. As reported in Table~\ref{tab:train-stats}, IEMOCAP exhibits higher entropy and lower inter-annotator agreement F-Kappa for human annotations when compared with that of MSP-Podcast, reflecting the greater ambiguity. However, synthetic annotations in IEMOCAP exhibit higher agreement and lower entropy than human labels, indicating reduced variability. It suggests that for highly ambiguous speech, synthetic labels may oversimplify nuanced emotions, yield less reliable distributions that limit complementarity, and potentially hinder model learning when combined with human annotations. To further probe this effect, we analyze performance across low, medium, and high ambiguity levels, defined by tertiles of entropy range (Fig.~\ref{fig:ambiguity-analysis}). In both datasets, combined annotations perform comparably to, or even surpass, human-only annotations under low and medium ambiguity, but the advantage diminishes under high ambiguity. 

These results suggest that in datasets with lower ambiguity, synthetic annotations can complement sparse human labels to improve distributional emotion modeling, while human annotations remain indispensable for capturing subtle and highly ambiguous emotions.

\vspace{-7pt}
\subsection{Effectiveness of DiME-Aug}
Table~\ref{tab:dime-results} shows that DiME-Aug improves performance across most configurations on both IEMOCAP and MSP-Podcast. On MSP-Podcast in particular, combined annotations achieve the best results after augmentation, confirming the effectiveness of DiME-Aug in mitigating imbalance, improving generalization, and producing more reliable distributional modeling. The only exception occurs with synthetic-only annotations on MSP-Podcast, where performance slightly declines after augmentation, likely because synthetic annotations already provide strong distributional approximations and further augmentation introduces redundancy or noise. 
\vspace{-5pt}
\section{Conclusion}
\label{sec:conclusion}
\vspace{-5pt}
This study investigated whether synthetic annotations generated by ALMs can complement sparse human labels to construct reliable emotion distribution representation. To this end, we proposed a three-component framework with Synthetic Perceptual Proxies, DiME-Aug, and ALM finetuning. 
Experiments show that synthetic annotations approximate human distributions with 6–10 samples per utterance in test datasets and complement human labels in naturalistic, low-ambiguity datasets, while human annotations remain essential for highly ambiguous emotions. Limitations include reliance on specific models and reduced effectiveness on acted emotions in IEMOCAP. Overall, our study suggests that ALM-generated synthetic annotations hold promise for advancing AER and supporting empathetic applications in domains like HCI, while it also highlight that achieving consistent reliability, particularly for highly ambiguous emotions, will require the development of more sophisticated prompting and generation strategies.


\bibliographystyle{IEEEbib}
\bibliography{refs}

\end{document}